\documentclass[aps,twocolumn,superscriptaddress,floatfix,preprintnumbers,nofootinbib]{revtex4-1}
\usepackage[utf8]{inputenc}
\usepackage{amsmath}
\usepackage{amssymb}

\usepackage{enumerate}
\usepackage{amsmath}
\usepackage{tikz}
\usepackage[compat=1.1.0]{tikz-feynman}
\usepackage{feynmf}
\usepackage{slashed}
\usepackage{braket}
\usepackage[toc,page]{appendix}
\usepackage{url}
\usepackage{natbib}
\usepackage{graphicx}
\usepackage[pdftex,bookmarks,linktocpage,pdfpagelabels,plainpages=false,hyperfigures,linkcolor=blue,citecolor=blue]{hyperref}
\hypersetup{colorlinks=true}

\usepackage{mathrsfs,amssymb}  
\usepackage{cancel}
\usepackage[normalem]{ulem}

\begin{document}

\begin{flushright}
\preprint{MI-TH-2029} 
\preprint{nuhep-th/20-13}
\end{flushright}

\author{Vedran~Brdar}
\email{vedran.brdar@northwestern.edu}
\affiliation{
Fermi National Accelerator Laboratory, Batavia, IL, 60510, USA}
\affiliation{Northwestern University, Department of Physics \& Astronomy, 2145 Sheridan Road, Evanston, IL 60208, USA}

\author{Bhaskar~Dutta}
\email{dutta@tamu.edu}
\affiliation{Mitchell Institute for Fundamental Physics and Astronomy,
Department of Physics and Astronomy, Texas A\&M University, College Station, TX 77843, USA}

\author{Wooyoung~Jang}
\email{wooyoung.jang@uta.edu}
\affiliation{Department of Physics, University of Texas, Arlington, TX 76019, USA}

\author{Doojin~Kim}
\email{doojin.kim@tamu.edu}
\affiliation{Mitchell Institute for Fundamental Physics and Astronomy,
Department of Physics and Astronomy, Texas A\&M University, College Station, TX 77843, USA}

\author{Ian~M.~Shoemaker}
\email{shoemaker@vt.edu}
\affiliation{Center for Neutrino Physics, Department of Physics, Virginia Tech, Blacksburg, VA 24061, USA}

\author{Zahra~Tabrizi}
\email{ztabrizi@vt.edu}
\affiliation{Center for Neutrino Physics, Department of Physics, Virginia Tech, Blacksburg, VA 24061, USA}

\author{Adrian~Thompson}
\email{thompson@tamu.edu}
\affiliation{Mitchell Institute for Fundamental Physics and Astronomy,
   Department of Physics and Astronomy, Texas A\&M University, College Station, TX 77843, USA}

\author{Jaehoon~Yu}
\email{jaehoon@uta.edu}
\affiliation{Department of Physics, University of Texas, Arlington, TX 76019, USA}

\title{Axion-like Particles at  Future Neutrino Experiments:\\
Closing the ``Cosmological Triangle"
}

\begin{abstract}
Axion-like particles (ALPs) provide a promising direction in the search for new physics, while a wide range of models incorporate ALPs. 
We point out that  future neutrino experiments, such as DUNE, possess competitive sensitivity to ALP signals. The high-intensity proton beam impinging on a target can not only produce copious amounts of neutrinos, but also cascade photons that are created from charged particle showers stopping in the target. 
Therefore, ALPs interacting with photons can be produced (often energetically) with high intensity via the Primakoff effect and then leave their signatures at the near detector through the inverse Primakoff scattering or decays to a photon pair. 
Moreover, the high-capability near detectors allow for discrimination between ALP signals and potential backgrounds, improving the signal sensitivity further. 
We demonstrate that a DUNE-like detector can explore a wide range of parameter space in ALP-photon coupling $g_{a\gamma}$ vs ALP mass $m_a$, including some regions unconstrained by existing bounds; the ``cosmological triangle'' will be fully explored and the sensitivity limits would reach up to $m_a\sim3-4$~GeV and down to $g_{a\gamma}\sim 10^{-8}~{\rm GeV}^{-1}$.
\end{abstract}

\maketitle

\noindent {\bf Introduction.} 
Axions not only address the strong CP problem \cite{Peccei:1977hh,Wilczek:1977pj,Weinberg:1977ma}, but also provide an explanation for the 27\% of the Universe's energy content constituting dark matter~\cite{Preskill:1982cy,Abbott:1982af,Dine:1982ah,Duffy:2009ig,Marsh:2015xka,Battaglieri:2017aum}. Investigation of the QCD axion has been extended to incorporate general axion-like particles (ALPs) in a wide range of models.
The experimental effort in the search for ALPs in these many incarnations is vigorously made using their couplings with Standard Model (SM) particles, primarily to photons, electrons, and nucleons.  
These experiments include helioscopes: CAST~\cite{Zioutas:1998cc,Anastassopoulos:2017ftl,Irastorza:2013dav}, haloscopes: Abracadabra~\cite{Kahn:2016aff,Salemi:2019xgl}, ADMX~\cite{Asztalos:2001tf,Du:2018uak}, CASPEr \cite{JacksonKimball:2017elr}, HAYSTAC \cite{Brubaker:2016ktl,Droster:2019fur}, light-shining-through-wall experiments: ALPSII~\cite{Spector:2019ooq}, interferometry \cite{Melissinos:2008vn,DeRocco:2018jwe}: ADBC~\cite{Liu:2018icu}, DANCE~\cite{Obata:2018vvr}, current and proposed 
accelerator-based experiments: FASER \cite{Feng:2018noy}, LDMX \cite{Berlin:2018bsc,Akesson:2018vlm}, NA62 \cite{Volpe:2019nzt}, NA64~\cite{Dusaev:2020gxi,Banerjee:2020fue}, SeaQuest~\cite{Berlin:2018pwi},  SHiP \cite{Alekhin:2015byh}, hybrids of beam dump and helioscope approaches: PASSAT~\cite{Bonivento:2019sri}, reactor experiments: MINER, CONUS etc.~\cite{Dent:2019ueq,AristizabalSierra:2020rom}, dark matter experiments: XENON~\cite{Aprile:2020tmw,Dent:2020jhf}, SuperCDMS~\cite{PhysRevD.101.052008}, PandaX~\cite{Fu_2017} etc.

The coupling to photons is particularly interesting, as it allows the ALPs to be copiously produced in accelerator-based neutrino experiments where, along with the neutrino flux, a high-intensity photon flux is generated from intense proton beams impinging on a target.
Photons emerge from bremsstrahlung and meson decays, and can convert to ALPs via the Primakoff process in the forward region.
The produced ALPs would then be observed at a detector via the inverse Primakoff scattering process, or via decays into photon pairs. A complementary analysis exploiting the axion-gluon coupling for QCD axions is done for the DUNE near detector~\cite{Kelly:2020dda}. 

In this Letter, we present the sensitivity of the ALP search at future neutrino experiments utilizing their near-detector facilities, taking DUNE as a concrete example, while the overall search strategies discussed here are readily applicable to other similar experiments.
The near-detector complex at DUNE will include three detectors, among which we will employ liquid argon (LAr) and gaseous argon (GAr) detectors for constraining parameter space of ALPs.

The DUNE near detectors will be located 574~m downstream of the target hall where the 120-GeV proton beam impinges on a a $1.5$-m long, segmented, cylindrical graphite target with a $16$-mm diameter.
The photon flux will be created with much higher energy compared to many ongoing accelerator-based neutrino experiments, e.g., COHERENT, CCM, JSNS$^2$ etc.~where $\mathcal{O}$(GeV) proton beams are used. 
The higher energy photons would allow us to probe a considerable range of ALP parameter space. 
We demonstrate that a DUNE-like experiment can explore a wide range of the ALP mass [from $\mathcal{O}(1)$ GeV to the massless limit] vs the ALP-photon coupling parameter space, covering some regions where there exist no (laboratory-based) constraints, e.g., the ``cosmological triangle'' and the region beyond the current beam-dump limits. 
An accurate estimate in the photon flux is essential for a more accurate estimate in the signal sensitivity. For this purpose, we simulate the photon flux using the \texttt{GEANT4}~\cite{Agostinelli:2002hh} package 
as conventional event generators are unable to handle secondary photon production such as cascade photons whose contribution is significant~\cite{Dutta:2020vop}.
We also consider the SM background to estimate more realistic constraints on the parameter space. 

\medskip 

\noindent {\bf ALP production at the target.} 
In order to investigate the ALP parameter space, we will focus on a generic model where the ALP field (henceforth denoted by $a$) can couple to a photon as described by interaction terms in the Lagrangian of the form
\begin{equation}
    \mathcal{L} \supset -\dfrac{1}{4}g_{a\gamma}  a F_{\mu\nu}\tilde{F}^{\mu\nu}\,,
\end{equation}
where $F_{\mu\nu}$ and $\tilde{F}_{\mu\nu}$ are the usual field strength tensor of the SM photon and its dual and where $g_{a\gamma}$ parameterizes the ALP-photon coupling in the unit of inverse energy. 
While this coupling appears at the tree level in DFSZ-type models of the QCD axion, it also arises in models of ALP dark matter and shows up naturally in string axiverse scenarios.

\begin{figure}[t]
 \centering
%
%
%
\includegraphics[width=8cm]{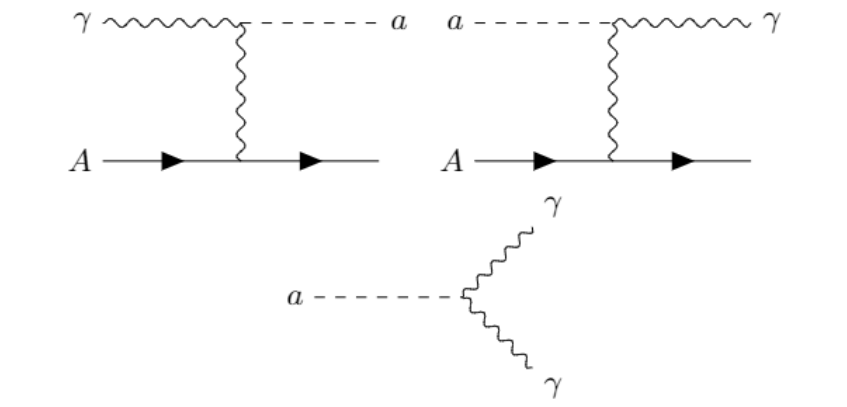}
\caption{Tree-level ALP production through the Primakoff process (top left), ALP scattering through the inverse Primakoff process (top right), and ALP decays (bottom).}
    \label{fig:compton_production}
\end{figure}
Given the above coupling, ALPs can be produced by the Primakoff scattering process, $\gamma+A\to a+A$ with $A$ symbolizing the atomic system of interest, as diagramatically displayed in Fig.~\ref{fig:compton_production}.
The production cross-section $\sigma_P$ in ALP scattering angle $d\Omega_a^\prime = \sin\theta_a^\prime d\theta_a^\prime d\phi_a^\prime$ is given by
\begin{equation}
    \frac{\partial^2\sigma_P}{\partial\theta_a^\prime \partial\phi_a^\prime} = \frac{g_{a\gamma}^2\alpha}{8\pi} \frac{p_a^4}{q^4} \sin^3\theta_a^\prime F^2(q) \,, \label{eq:xsinangle}
\end{equation}
where $\alpha$ and $m_a$ are the fine structure constant and the ALP mass, respectively and where $q^2=m_a^2-2E_\gamma(E_a-p_a\cos\theta_a^\prime)$. Here, form factor $F^2(q)$ encapsulates an important coherent enhancement of the atomic number $Z^2$.

For the high energy photon flux in the target, the relevant coherence length (and momentum transfer) will be at the nuclear scale. We therefore use the nuclear form factor, for which we adopt the Helm parameterization
\begin{equation}
    F_N^2(q) = Z^2\bigg(\dfrac{3 j_1 (qR_1)}{qR_1}\bigg)^2 e^{-q^2 s^2}\,,
\end{equation}
with $s=0.9$~fm and $R_1=\sqrt{(1.23A^{1/3}-0.6)^2+2.18}$~fm with $A$ being the atomic mass number. 
The results do not change if we replace the nuclear form factor by the atomic form factor.

We next discuss how the ALP flux is connected to the photon flux at DUNE. 
Let us work in the frame $(x,y,z)$ such that $z$ points along the beam axis and each photon created in the target lies in the $xz$ plane with an angle $\theta_\gamma$ with respect to the $z$ axis. Then, the conversion event on an atom at $(0,0,0)$ involves a photon with unit vector $\hat{\gamma}=(\sin\theta_\gamma, 0, \cos\theta_\gamma)$ and an ALP, which is generated at an angle $\theta_a^\prime$ with respect to the photon direction according to the angular distribution generated by $\partial^2\sigma_P / \partial\theta_a^\prime \partial\phi_a^\prime$ as in Eq.~\eqref{eq:xsinangle}. 
The ALP direction in the frame $(x^\prime, y^\prime, z^\prime)$ where the photon momentum points along the $z^\prime$ axis is
\begin{align}
    \hat{a}^\prime &= (\cos\phi_a^\prime \sin\theta_a^\prime, \sin\phi_a^\prime \sin\theta_a^\prime, \cos\theta_a^\prime)\,
\end{align}
where $\phi_a^\prime$ is an azimuthal angle around the $z'$ axis.
$\hat{a}'$ can be transformed to $\hat{a}$ defined in the unprimed coordinate frame by a rotation by $-\theta_\gamma$ about the $y'$ axis:
\begin{align}
    \hat{a} = (&\cos\theta_\gamma \cos\phi_a^\prime \sin\theta_a^\prime - \cos\theta_a^\prime \sin\theta_\gamma, \sin\theta_a^\prime \sin\phi_a^\prime, \nonumber \\
    &\cos\theta_a^\prime \cos\theta_\gamma + \cos\phi_a^\prime \sin\theta_a^\prime \sin\theta_\gamma)\,,
\end{align}
from which we find that the polar angle of ALP $\theta_a$ measured in the unprimed frame is
\begin{equation}
    \theta_a = \arccos (\cos\theta_a^\prime \cos\theta_\gamma + \cos\phi_a^\prime \sin\theta_a^\prime \sin\theta_\gamma)\,.
    \label{eq:angle}
\end{equation}

Assuming that the face of the detector spans a circular aperture of radius $r$, a distance $\ell$ from the target for simplicity, we see that the ALP flux would enter the detector as far as $\theta_a$ is less than the detector opening angle $\theta_{\rm det}=\arctan(r/\ell)$.
We then convolve the differential photon flux in $E_\gamma$ and $\theta_\gamma$ with the differential Primakoff production cross-section in Eq.~\eqref{eq:xsinangle}:
\begin{eqnarray}
    \frac{d\Phi_a}{dE_a} = && \hspace{-0.4cm}\int \frac{\partial^2 \Phi_\gamma}{\partial E_\gamma \partial \theta_\gamma} \bigg[\dfrac{1}{\sigma_P + \sigma_\gamma} \frac{\partial^2\sigma_P}{\partial\theta_a^\prime \partial\phi_a^\prime}\bigg] \delta(E_a - E_\gamma) \nonumber \\ 
    & \times& \Theta(\theta_\text{det} - \theta_a) d\phi_a^\prime d\theta_a^\prime d\theta_\gamma \,.
    \label{eq:alp_flux_1}
\end{eqnarray}
The factor of $\frac{1}{\sigma_P + \sigma_\gamma} \frac{\partial^2\sigma_P}{\partial\theta_a^\prime \partial\phi_a^\prime}$ is the differential branching fraction of ALP production, which is used to take into account the fraction of photons that convert into ALPs in the target versus those that are absorbed through standard interactions such as photoelectric absorption and pair production. One may find that above an MeV, the total photon absorption cross-section $\sigma_\gamma \sim 1$~barn in carbon.

Finally, we discuss the generation of photon flux before closing this section. 
A large fraction of photons are produced via the decays of mesons such as $\pi^0$ and $\eta$ and radiation off the incoming beam. These contributions can be estimated using conventional event generators. 
There are additional significant contributions, e.g., cascade photons from secondary particles such as ionized electrons created while final-state particles lose their energy and stop in the target~\cite{Dutta:2020vop}. 
The estimate of those photons involves non-trivial nuclear effects, so we employ the \texttt{GEANT4} package for a systematic estimate of the photon flux, simply considering the target geometry and the beam specifications of a DUNE-like experiment with a 1.2-MW, 120-GeV proton beam impinging on a 
graphite target. We assumed a proton-on-target (POT) rate of $1.1 \times 10^{21}$ POT$\cdot$year$^{-1}$.
We find that most of the photons are moving in the forward direction, hence one can expect that a large fraction of ALP flux is directed toward the near detector complex. 

\medskip

\noindent {\bf ALP event rate at the detector.}
ALPs can be detected by inverse Primakoff scattering or decays to $\gamma\gamma$ at the detector. 

For the decay to a photon pair, we compute the probability $P_{\rm decay}$ that the ALP decays within the detector volume. 
This can be calculated by integrating the decay probability density between the front and the back of the detector, $(\ell, \ell+\Delta\ell)$;
\begin{equation}
    P_{\rm decay}= e^{-\ell/(\tau v_a)} \left[ 1 - e^{-\Delta\ell /(\tau v_a) } \right]\,,
\end{equation}
where $\tau$ is the ALP lifetime in the laboratory frame, based on the decay width
\begin{equation}
    \Gamma (a\to\gamma\gamma) = \dfrac{g_{a\gamma}^2 m_a^3}{64 \pi},
\end{equation}
and velocity in the laboratory frame $v_a = p_a / E_a$. 
For a given exposure time $\mathcal{E}$, the total event rate from ALP decays $N_{\rm decay}$ is then given by a convolution of the decay probability with the ALP flux derived in Eq.~\eqref{eq:alp_flux_1}:
\begin{equation}
    N_{\rm decay} = \mathcal{E}\int \frac{d\Phi_a}{dE_a}P_{\rm decay} dE_a\,.
\end{equation}

Similarly, ALPs can scatter, yielding a single $\gamma$ signal, through the inverse Primakoff scattering process. This process has a cross-section $\sigma_{IP}$, identical to $\sigma_P$ but larger by a factor of 2 to account for the sum over polarization states.
\begin{equation}
    N_{\rm scatter} = N_T \mathcal{E} \int \sigma_{IP}(E_a)\frac{d\Phi_a}{dE_a} P_{\rm surv} dE_a\,,
\end{equation}
where $N_T$ is the number of argon targets. 
$P_{\rm surv}$ is the survival probability that an ALP reaches the detector without decaying, that is, $P_{\rm surv}=e^{-\ell /(\tau v_a)}$. 

\medskip

\begin{figure}[t]
    \centering
    \includegraphics[width=0.5\textwidth]{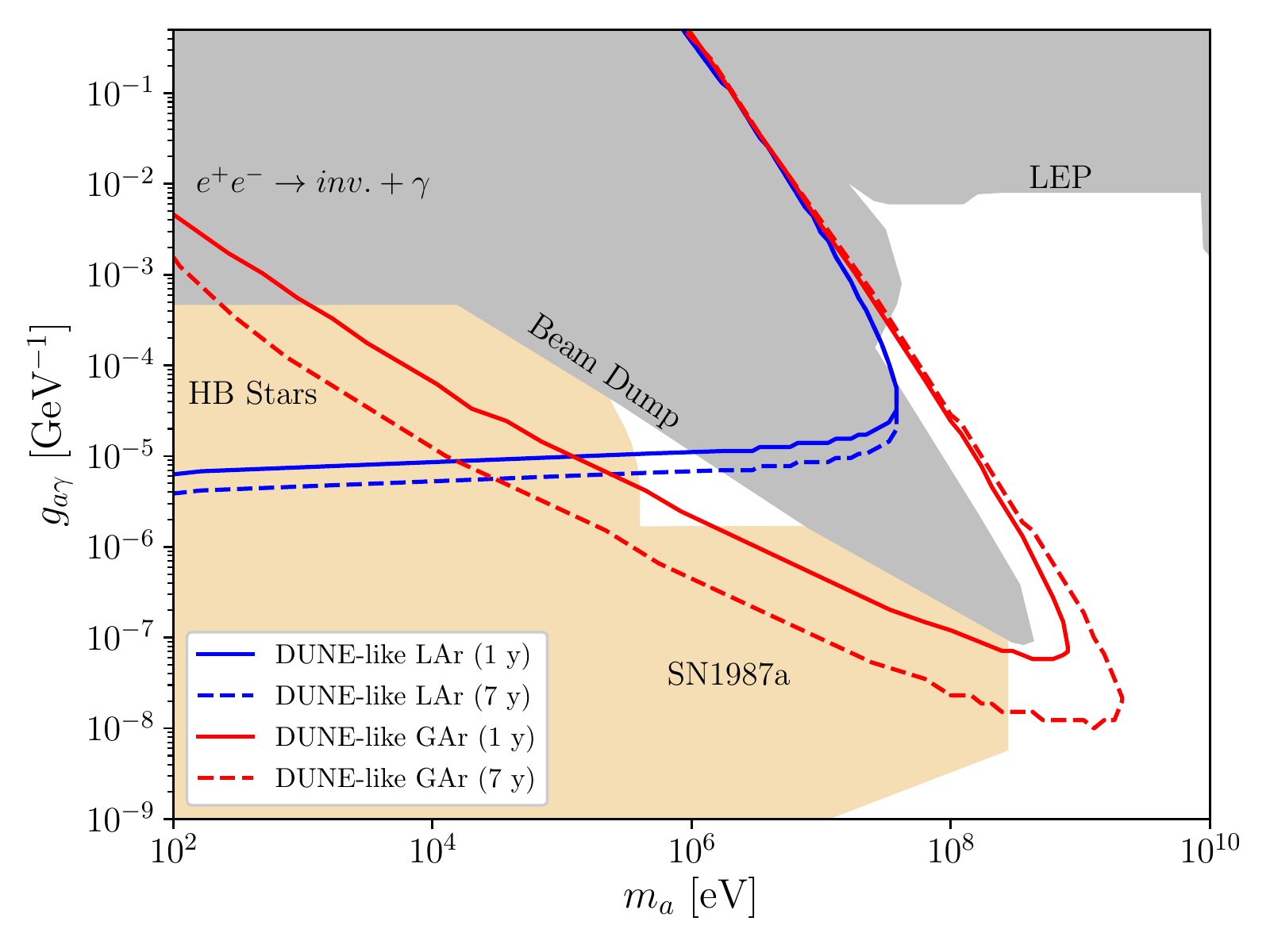}
   \caption{The 90\% C.L. sensitivity reaches in the $(m_a, g_{a\gamma})$ plane expected for a DUNE-like GAr detector with 1 and 7 year exposures ($a\to\gamma\gamma$ decays) and at a LAr detector for 1 and 7 year exposures (inverse Primakoff scattering, $a + A \to \gamma + A$). Existing laboratory limits~\cite{Blumlein:1991xh,Blumlein:1990ay,Jaeckel:2015jla} are shown in gray, while astrophysical limits~\cite{Raffelt:1985nk,Raffelt:1987yu,Raffelt:2006cw,Payez:2014xsa,Jaeckel:2017tud} are shown in tan.}
    \label{fig:limits}
\end{figure}

\noindent {\bf Background consideration.}
In Fig.~\ref{fig:limits}, one can infer two qualitatively different exclusion regions corresponding to scatterings and decays of ALPs. The limits will be discussed in more detail in the next section and here we note that they were obtained under a zero-background hypothesis. In what follows, we justify such a choice for the case of decaying ALPs in a DUNE-like GAr detector and also discuss the situation in a DUNE-like LAr detector for scattering ALPs.
For a DUNE-like GAr detector, we make use of its volume and consider ALP decays into two photons. The main SM background is from NC-$\pi^0$ production where the final-state pion decays also to two photons.

For the 7-year exposure, the DUNE Collaboration predicts $\sim3\times 10^6$ NC events in the detector~\cite{ichep}, summing up the contributions from positive and negative horn polarity modes.
Let us conservatively assume that all of those would contain a single pion. Among those, the events where there is some hadronic activity in the interaction vertex could be vetoed as ALPs decay would not give rise to additional energetic charged particles. 
The study in \cite{Berryman:2019dme} estimated that the cut on the hadronic activity removes $\sim 80\%$ of the background. 

\begin{figure*}
    \centering
    \includegraphics[width=0.3\textwidth]{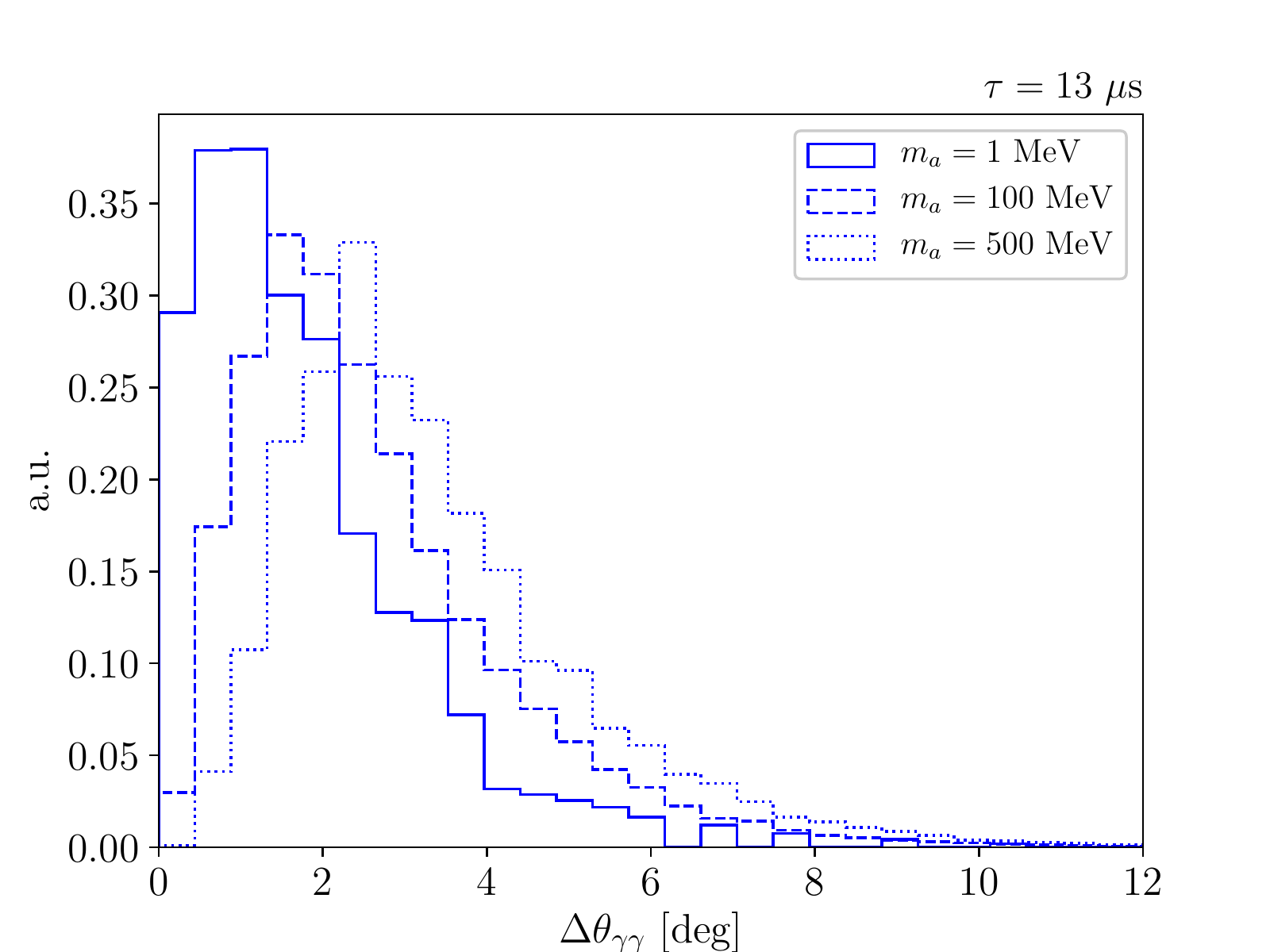}
    \includegraphics[width=0.3\textwidth]{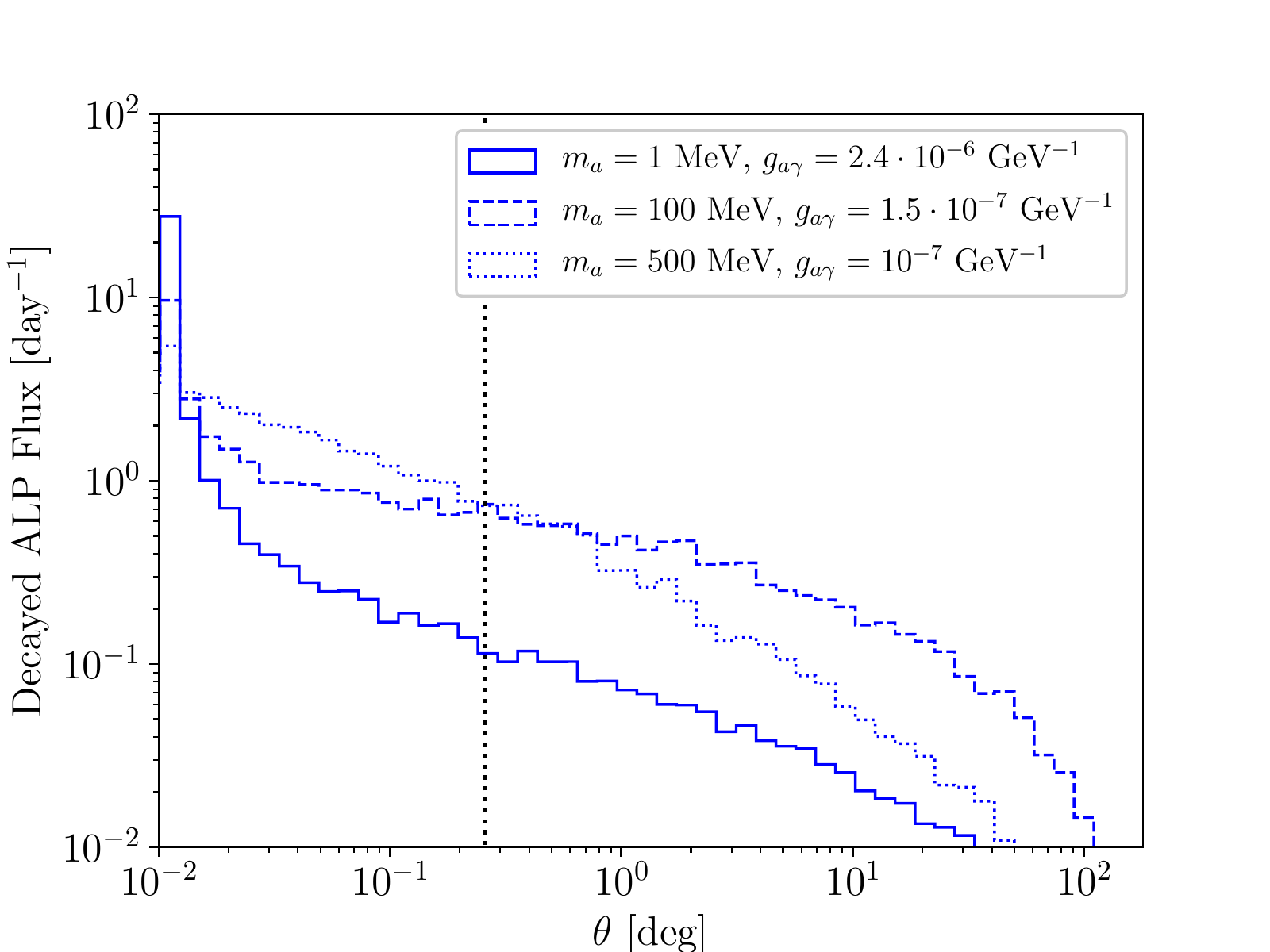}
    \includegraphics[width=0.3\textwidth]{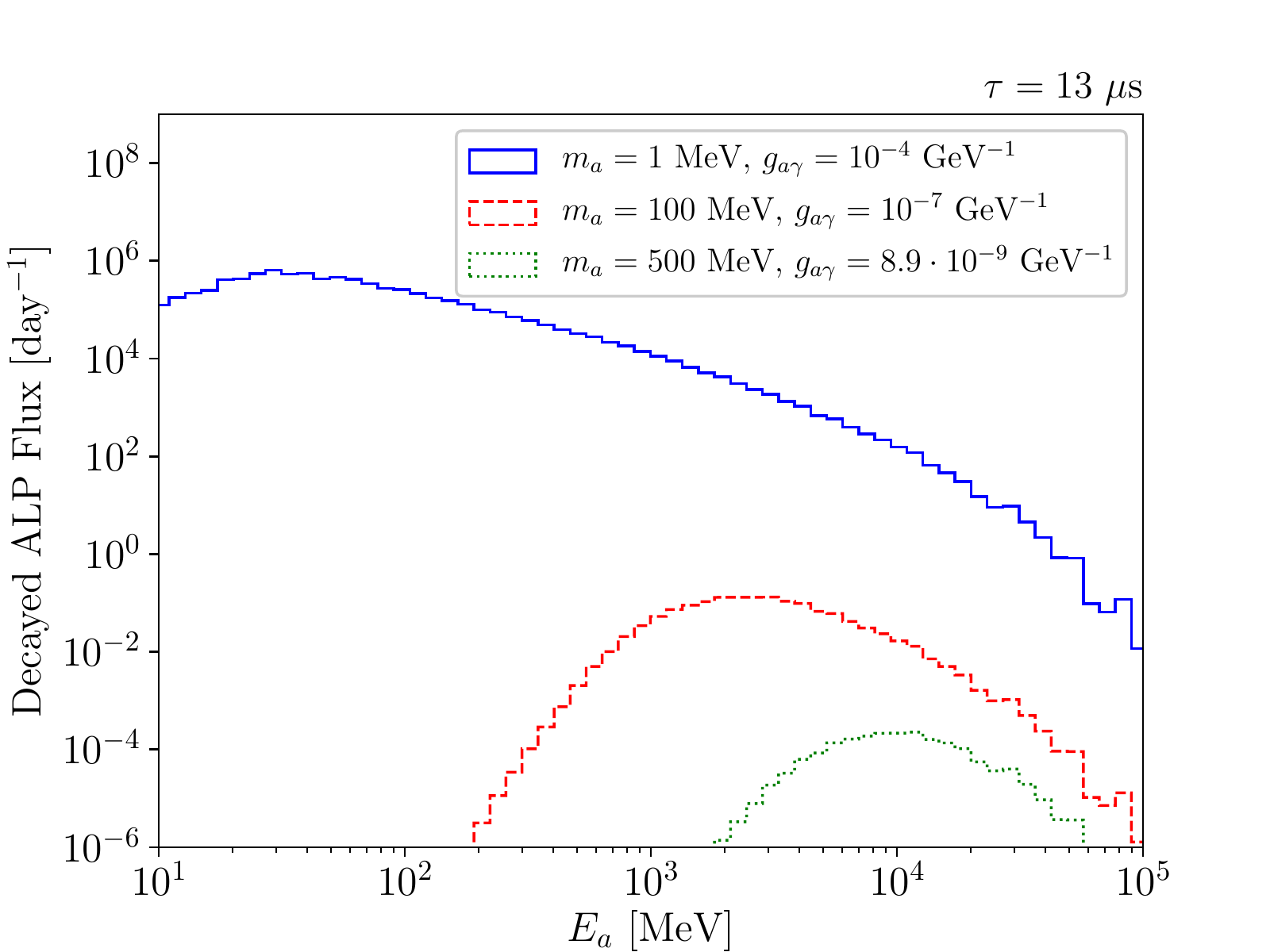}
    \caption{Left: Opening angle $\Delta\theta_{\gamma\gamma}$ from $a\to\gamma\gamma$ decays for representative ALP masses in the decay-dominated signal regime for a fixed lifetime. Middle: Decaying ALP fluxes as a function of the angle with respect to the beam axis. The black dotted line represents the geometric acceptance for the GAr-detector solid angle. Right: Normalized energy spectra for the decay-dominated ($a\to\gamma\gamma$) region of parameter space for a fixed ALP lifetime.}
    \label{fig:opening_angle}
\end{figure*}

For a further background suppression, one can employ kinematics: while the ALPs and hence its decay products (photons) will appear in the detector at a very small opening angle (see also the left panel of Fig.~\ref{fig:opening_angle}) and are practically collimated with the beamline, the $\pi^0$s in the detector (and hence photons from their decay) would be distributed with a greater preference than our signal toward larger angles. We have checked that explicitly, using Monte Carlo generator \texttt{GENIE}~\cite{andreopoulos2015genie}. We found that by imposing a cut on the angle $\epsilon=4$ mrad, which matches resolution of the detector \cite{Berryman:2019dme}, background is reduced by the factor of $10^3$.

For the coherent single photon production (not yet implemented in \texttt{GENIE}), we expect negligible contributions, considering its 3-4 orders of magnitude smaller cross-sections than those of NC-$\pi^0$ events~\cite{Rein:1981ys,Wang:2013wva}. For completeness, let us emphasize that the background from misidentified electrons is suppressed; the $\nu_e$ flux producing electrons in CC processes is suppressed with respect to the $\nu_\mu$ flux. Further, the dirt events which could be present in both LAr and GAr analyses are subdominant with respect to the in-detector $\pi^0$ background; this is mainly because to generate dirt events, neutrinos would need to interact very close to the upstream end of the LAr detector and this dramatically reduces effective volume of the dirt. The few events that may sneak into the detector can be removed using topological and spatial cuts. See the related discussion for MiniBooNE~\cite{Aguilar-Arevalo:2020nvw}.

Finally, the likelihood for a photon conversion in the gas 
is only 12\%, which means that there is 23\% chance that at least one of the photons would convert. This reduces the background counts by a further factor of 4, but we note that our signal also suffers from this. Putting all the numbers together leaves us with $\mathcal{O}(10)$ background events. We note that this number could be further reduced by employing kinematic variables (which depend on $m_a$). 
For example, it is possible to use the opening angle, $\Delta\theta_{\gamma\gamma}$, between the photons (the left panel of Fig.~\ref{fig:opening_angle}) and the photon energy spectra (the right panel of Fig.~\ref{fig:opening_angle}) emerging from the ALP decays for various $m_a$ to suppress the backgrounds.
Still, note that by even taking into account background of this level would basically not yield an observable change of the presented limits.

As far as the limit from ALP scattering in LAr is concerned, here the situation is vastly different.
First, the fiducial mass of LAr detector is 50 times larger than that of GAr, hence giving $\mathcal{O}(10^8)$ $\pi^0$ events before cuts. We anticipate that most of those would typically yield two observable showers  and hence would be rejected since, in contrast, there would be only one final-state photon from the ALP scattering.
For eliminating the remaining $\pi^0$ events where one of the photons would be missed, one can employ kinematical distributions.
Regarding those, unlike in the decaying case, one cannot assume that all the photons from ALPs will be within an $\epsilon$ angle because photons emerging from ALP scattering in the detector would have distribution similar to those arising from decaying $\pi^0$. This makes the background consideration more involved than in the case of decaying ALPs. We reserve a more quantitative study in this regard for a follow-up work and present a background-free limit in what follows.

\medskip 

\noindent {\bf Results.}
We are now in the position to discuss the expected sensitivity of a DUNE-like detector to the ALP scattering and decay signals in the presence of SM backgrounds. 
 We consider the LAr detector of a 50-ton fiducial mass for the scattering limit and the GAr detector of a cylindrical fiducial volume (5.2-m diameter and 5-m length) for the decay limit.
Our sensitivity reach for decays are then evaluated at 90\% C.L. and reported in Fig.~\ref{fig:limits} by the red lines, while the reach using inverse Primakoff scattering is shown in blue.
Various existing astrophysical and laboratory-based limits are also shown by tan and gray regions, respectively. For lower ALP masses, notice that the scattering limit becomes much better than the decay limit near $m_a \sim 100$ keV. 

This study suggests that a DUNE-like detector, especially, the gas-phase detector due to its significantly low background contamination, will be greatly capable of probing a wide range of the $(m_a,g_{a\gamma})$ parameter space. A few remarkable potentials are worth pointing out.
Above all, a DUNE-like detector can probe the regions beyond the current beam-dump limits, i.e., up to $m_a\sim 3-4$~GeV and down to $g_{a\gamma}\sim10^{-8}~{\rm GeV}^{-1}$, which have never been explored before by existing direct or indirect searches. 
Second, the cosmological triangle surrounded by beam-dump, HB stars, and supernova limits can be (completely) covered. Although this region would be constrained by standard cosmological considerations, they are highly model-dependent, hence can be evaded by non-standard cosmology~\cite{Carenza:2020zil,Depta:2020wmr}. 
Third, a large portion of astrophysical limits (HB stars and SN1987a) can be constrained (see also \cite{Lucente:2020whw} for a recent development in the supernova limit calculation and \cite{Bar:2019ifz} for its criticism). Like the cosmological limits, the astrophysical limits highly depend on the underlying ALP model details (see, for example, Refs.~\cite{Jaeckel:2006xm,Khoury:2003aq,Masso:2005ym,Masso:2006gc,Dupays:2006dp, Mohapatra:2006pv,Brax:2007ak,DeRocco:2020xdt}), so a DUNE-like detector is expected to constrain those regions in a model-independent fashion.  A number of ongoing/upcoming  experiments, e.g., Belle II and  neutrino experiments having smaller beam energy (e.g., CCM, SBND, T2K/T2HK) would also be able to investigate the cosmological triangle using the similar production process as described here but DUNE is going to have much better reach for smaller $g_{a\gamma}$/larger $m_a$ values. 

\medskip 

\noindent {\bf Outlook.}
We plan to extend our analysis in several directions in future work. For example, DUNE-PRISM may offer advantages for further background discrimination given its ability to move several degrees off-axis. We have also not considered a detailed background estimation for the scattering regime relevant for ALPs below 0.1 MeV. Lastly, we expect other neutrino facilities 
to play a complementary role in the future hunt for ALPs.   
     
\acknowledgements
We would like to thank Alan Bross,  Andr\'{e}~de~Gouv\^{e}a,  Laura Fields, Vladimir Ivantchenko, Soon Yung Jun and Shirley Li for useful discussions. We also thank Patrick Huber to be part of this work in the initial phase. JY also acknowledges discussions with several other members of the DUNE Collaboration.
BD and AT acknowledge support from the U.S. Department of Energy (DOE) Grant DE-SC0010813.
The work of DK is supported by DOE under Grant No. DE-FG02-13ER41976/DE-SC0009913/DE-SC0010813.  
The work of IMS is supported by DOE under the award number DE-SC0020250. 
The work of ZT is supported by DOE under the award numbers DE-SC0020250 and DE-SC0020262.
WJ and JY acknowledge the support from DOE under Grant No. DE-SC0011686.
Fermilab is operated by the Fermi Research Alliance, LLC under contract No. DE-AC02-07CH11359 with the United States Department of Energy. Portions of this research were conducted with the advanced computing resources provided by Texas A\&M High Performance Research Computing.

\bibliography{main}

\end{document}